\documentclass[12pt,a4paper,twoside]{article}
\newenvironment{changemargin}[2]{%
\begin{list}{}{%
\setlength{\leftmargin}{#1}%
\setlength{\rightmargin}{#2}%
}%
\item[]}
{\end{list}}
\usepackage{graphicx}
\usepackage{color}
\usepackage{lscape}
\usepackage{hyperref}
\usepackage{epstopdf}
\usepackage{lscape}
\usepackage{amsmath}
\usepackage{hyperref}
\usepackage{amsmath}
\usepackage{setspace}
\begin{document}
\baselineskip=0.30in
{\bf \LARGE

\begin{changemargin}{-1.2cm}{0.5cm}
\begin{center}
Analysis of quantum-mechanical states of the Mie-type ring shaped  model via the Fisher's information entropy
\end{center}

\end{changemargin}
}
\vspace{4mm}
\begin{center}
{\Large{\bf B. J. Falaye $^a$$^{,}$$^\dag$$^{,}$}}\footnote{\scriptsize E-mail:~ fbjames11@physicist.net;~ babatunde.falaye@fulafia.edu.ng\\ \dag{Corresponding} author Tel.: +2348103950870}\Large{\bf ,} {\Large{\bf K. J. Oyewumi $^b$$^{,}$}}\footnote{\scriptsize E-mail:~ kjoyewumi66@unilorin.edu.ng}\Large{\bf ,} {\Large{\bf S. M. Ikhdair $^c$$^{,}$}}\footnote{\scriptsize E-mail:~ sikhdair@neu.edu.tr;~ sikhdair@gmail.com.} \Large{\bf and} {\Large{\bf M. Hamzavi $^d$$^{,}$}}\footnote{\scriptsize E-mail:~ majid.hamzavi@gmail.com}
\end{center}
{\small
\begin{center}
{\it $^\textbf{a}$Applied Theoretical Physics Division, Department of Physics, Federal University Lafia,  P. M. B. 146, Lafia, Nigeria.}
{\it $^\textbf{b}$Theoretical Physics Section, Department of Physics, University of Ilorin,  P. M. B. 1515, Ilorin, Nigeria.}
{\it $^\textbf{c}$Department of Physics, Faculty of Science, an-Najah National University, New campus, P. O. Box 7, Nablus, West Bank, Palestine.}
{\it $^\textbf{d}$Department of Physics, University of Zanjan, Zanjan, Iran.}
\end{center}}

\begin{abstract}
\noindent
In the recent years, information theory of quantum-mechanical systems have aroused the interest of many Theoretical Physicist. This due to the fact that it provides a deeper insight into the internal structure of the systems. Also,  It is the strongest support of the modern quantum computation and information, which is basic for numerous technological developments. This study report the any $\ell-$state solution of the radial Schr\"{o}dinger equation with the Mie-type ring shaped diatomic molecular potential. Rotational-vibration of some few selected diatomic molecules are given. The probability distribution density of the system which gives the probability density for observing the electron in the state characterized by the quantum numbers $(n, l, m)$ in the Mie-type ring shaped diatomic molecular potential is obtained. Finally, we analyze this distribution via a complementary information measures of a probability distribution called as the Fisher's information entropy.
\end{abstract}


 diatomic molecular potential.

{\bf PACs No.}: 03.65.–w, 03.65.Ge, 03.67.–a; 03.70.+k; 03.65.-a.

\begin{changemargin}{-1.2cm}{0.5cm}
\section{Introduction}
\label{sec1}
In the recent years, noncentral potentials have received a lot of attention  in various fields of physics.  This is due to the occurrence of ``accidental'' degeneracy and ``hidden'' symmetry in those noncentral potentials. In quantum chemistry and nuclear physics, these noncentral potentials are use in the descriptions of ring-shaped molecules like benzene
and the interactions between deformed pair of nuclei. This study started with the pioneering work of Hartmann \cite{IN1,IN2,IN3} and Makarov et al \cite{IN4} on ring-shaped potentials.

Because of its applications in quantum chemistry and nuclear physics, numerous studies have been reported in the recent years. Quesne proposed a ring-shaped potential in which
the Coulomb part of the Hartmann potential was replaced by a harmonic oscillator term. In addition, the dynamical invariance algebra and ’accidental degeneracy’ have been
carried out for this new model potential \cite{IN5}. Recently, Chen and Dong found a new ring-shaped (noncentral) potential and obtained the exact solution
of the Schr\"{o}dinger for the Coulomb potential plus this new ring shaped potential \cite{IN6}. Shortly thereafter, Ikhdair and Sever \cite{IN7}, proposed a new non-central potential. They obtained the energy eigenvalues and eigenfunctions of the bound-states for the Schr\"{o}dinger equation in D-dimensions for the potential via the Nikiforov-Uvarov method. Cheng and Dai \cite{IN8}, proposed a new potential consisting of a modified Kratzer’s potential \cite{IN9} plus the new proposed ring-shaped potential in \cite{IN6}. 

Furthermore, the Schr\"{o}dinger equation with noncentral electric dipole ring-shaped potential have been investigated by working in a complete square integrable basis that supports an infinite tridiagonal matrix representation of the wave operator \cite{IN10}. Very recently, Falaye obtained the solutions of three dimensional Klein-Gordon equation for the spherically and non-spherically harmonic oscillatory ring-shaped potentials within the framework of asymptotic iteration method\cite{IN11}. 

These contributions were achieved by solving the Schr\"{o}dinger equation through separating the variables in spherical, parabolic or other orthogonal curvilinear coordinate systems. Few of the approaches which have been employed in solving this problems include the asymptotic iteration method (AIM) \cite{IN11}-\cite{F22}, Feynman integral formalism \cite{F23,F24,F25}, functional analysis approach \cite{F26, F27, F8, F29, F30, F31}, exact quantization rule method \cite{F32, F33, F34, F35, F36, F37}, proper quantization rule \cite{F38, F40}, Nikiforov-Uvarov (NU) method \cite{F41, F42, F43, F44, F45, F46, F47, F48, F49}, supersymetric quantum mechanics \cite{F50}-\cite{F65}, etc. However, the information-theoretic quantities remain to be computed. This may be due to  lack of knowledge in the information-theoretic properties of special functions. It is therefore the priority purpose of the present work to find the bound state solution of Mie-type ring shaped diatomic molecular potential and then to obtain the information-theoretic measures for this potential via the Fisher's information entropy.

Shannon entropy (S) and the Fisher's information measure (F) of the probability distributions are becoming increasingly important tools of scientific analysis in a variety of disciplines of scientific inquiry \cite{IN12, IN13}. For instance, the Frieden extreme physical information principle uses the Fisher's information measure to derive important laws of chemistry and physics, such as the equations of the nonrelativistic quantum mechanics or relevant results in density functional theory. Jaynes’ maximum entropy principle, which utilizes S, provides a method for constructing the whole of statistical thermodynamics which has led to a large variety of applications centered around S. There has been several worthy attempt in applying the two probability distribution functions in physics and related areas \cite{IN12,DG1,IN13,IN14,IN15,IN16,IN17,IN18}. These implies that both S and F can be used as complementary tools to describe the information behavior, pattern, or complexity of physical systems and the electronic processes involving them \cite{IN12}.

The F controls the localization of the density around its nodes \cite{IN19}. In this sense, it is also a measure of the oscillatory degree of the wavefunction. Further, the F
has been shown to be closely related to some density functionals describing physical and chemical observables \cite{IN19,IN20}. In particular, it describes the
Weisz\"acker energy functional of atomic and molecular systems \cite{IN19,IN20,IN21} and it has an intimate connection with the kinetic energy \cite{IN22}. Recently, Romera and Dehesa \cite{IN23} used the F to analyze some correlation properties of many-electron systems, to characterize the so-called avoided crossings of atomic systems in external electric and magnetic fields, and to derive the uncertainty relation of Cramer–Rao type.

For rank-2 density matrices and operators with zero diagonal elements in the eigenbasis of the density matrix, Toth and Petz prove analytically that the quantum Fisher's information is four times the convex roof of the variance. Strong numerical evidence suggests that this statement is true even for operators with nonzero diagonal elements or density matrices with a rank larger than 2 \cite{IN24}. Polettini and Esposito describe a general method based on the Fisher's information matrix to discriminate between generators that do and don't admit nonconvex solutions. The initial conditions leading to concave transients are shown to be extremely fine-tuned, by their method they are able to select nonconvex initial conditions that are arbitrarily close to the steady state \cite{IN25}.

The schematic presentation of our research is as follows: In the next section, we obtain the any $\ell-$state solution of the radial Schr\"{o}dinger equation with the Mie-type ring shaped diatomic molecular potential. The probability distribution density of the system is also calculated. In section 3, The information-theoretic measures for the Mie-type ring shaped diatomic molecular potential is presented. Numerical discussion are given in section 4 and we give brief conclusion in section 5.
\section{Any $\ell-$state solution of the radial Schr\"{o}dinger equation with the Mie-type ring shaped diatomic molecular potential}
\label{sec2}
The ring shaped Mie-type potential which we examine in this study can be defined as:
\begin{equation}
V(r,\theta)=D_e\left[\frac{k}{j-k}\left(\frac{r_e}{r}\right)^j-\frac{j}{j-k}\left(\frac{r_e}{r}\right)^k\right]+\eta\frac{\cos^2\theta}{r^2\sin^2\theta},
\label{E1}
\end{equation}
where the parameter $D_{e}$ determines the interaction energy between two atoms in a solid at $r=r_{e}$, $r_e$ is the molecular bond length and $\eta$ is positive real constants. We find that this potential reduces to the Mie-type potential in the limiting case of $\eta=0$.  By taking $j=2k$ and further setting $k=1$, the potential reduces to the ring shaped Kratzer-Fues potential which is formed from combination of Kratzer-Fues \cite{BJ1, BJ2, BJ3, BJ4, BJ5} and the angular potential. On the other hand, another form of the potential, called as the ring modified Kratzer potential can be obtained by adding parameter $D_e$ to the radial part of the formal. Since our interest is in these two potentials\footnote{ring shaped Kratzer-Fues potential and ring shaped modified Kratzer potential}, we therefore consider a general form of the potential as:
\begin{equation}
V(r,\theta)=\frac{a}{r^2}-\frac{b}{r}+c+\eta\frac{\cos^2\theta}{r^2\sin^2\theta},\ \ \ \ \ \  a=D_er_e^2,\ \ b = 2D_er_e,\ \ c = D_e.
\label{E2}
\end{equation}
The shape of the two potentials of interest are display in figures (\ref{fig1}) and (\ref{fig2}). Our aim in this section is to derive the energy spectrum for a moving charged particle in the presence of a potential given by the above equation (\ref{E2}) analytically in a very simple way. Thus, we begin by writing Schr\"{o}dinger equation in spherical polar coordinates for a particle in the presence of a potential $V(r,\theta)$ as:
\begin{equation}
\left(-\frac{\hbar^2}{2\mu}\left[\frac{1}{r^2}\frac{\partial}{\partial r}r^2\frac{\partial}{\partial r}+\frac{1}{r^2\sin \theta}\frac{\partial}{\partial\theta}\left(\sin\theta\frac{\partial}{\partial\theta}\right)+\frac{1}{r^2\sin^2\theta}\frac{\partial^2}{\partial\phi^2}\right]+V(r,\theta)\right)\Psi_{n\ell m}(r,\theta,\varphi)=E_{n\ell}\Psi_{n\ell m}(r,\theta,\varphi),
\label{E3}
\end{equation}
where $\mu=\frac{m_1m_2}{m_1+m_2}$ being the reduced mass, $n$ denotes the radial quantum number ($n$ and $\ell$ are named as the vibration-rotation quantum numbers in molecular chemistry), $r$ is the internuclear separation, $E_{n\ell}$ is the exact bound state energy eigenvalues, $V(r,\theta)$ is the internuclear potential energy function and $\Psi_{n\ell m}(r,\theta,\varphi)$ denotes the total wave function which is defined by $\Psi_{n\ell m}(r,\theta,\varphi)=R_{n\ell}(r)Y_{\ell m}(\theta,\phi)=R_{n\ell}(r)\Theta(\theta)e^{\pm im\varphi}$. On substituting this wave function into equation (\ref{E3}) and then considering potential (\ref{E1}), we obtain the following two ordinary differential equations; one for the radial part and the other for the angular part
\begin{subequations}
\begin{equation}
\frac{d^2R_{n\ell}(r)}{dr^2}+\frac{2\mu}{\hbar^2}\left[E+\frac{b}{r}-c-\frac{a+\ell(\ell+1)}{r^2}\right]R_{n\ell}(r)=0,
\label{E4a}
\end{equation}
\begin{equation}
\frac{d^2\Theta(\theta)}{d^2\theta}+\frac{\cos\theta}{\sin\theta}\frac{d\Theta(\theta)}{d\theta}+\left[\ell(\ell+1)-\frac{m^2+\eta\cos^2\theta}{\sin^2\theta}\right]\Theta(\theta)=0,
\label{E4b}
\end{equation}
\end{subequations}
respectively. For $\Psi_{n\ell m}(r,\theta,\varphi)$ to be finite everywhere, $R_{n\ell}(r)$ must vanish at r = 0, that is, $R_{n\ell}(r)=0$, then $R_{n\ell}(r)$ is a real function. Therefore, the solution for the radial part of the equation can be easily found in terms of degenerate hypergeometric function as
\begin{equation}
R_{n\ell}(r)=N_{n\ell}\ r^{-\frac{1}{2}+\sqrt{\left(\ell+\frac{1}{2}\right)^2+\frac{2\mu a}{\hbar^2}}} \exp\left(-\sqrt{\frac{2\mu}{\hbar^2}(c-E_{n\ell})} r\right)L_n^{\sqrt{\left(\ell+\frac{1}{2}\right)^2+\frac{2\mu a}{\hbar^2}}}\left(2\sqrt{\frac{2\mu}{\hbar^2}(c-E_{n\ell})} r\right).
\label{E5}
\end{equation}
and the desired bound-state satisfying the boundary conditions can also be found as
\begin{equation}
E_{n\ell}=c-\frac{2\mu b^2}{\hbar^2}\left[1+2n+2\sqrt{\left(\ell+\frac{1}{2}\right)^2+\frac{2\mu a}{\hbar^2}}\right]^{-2}
\label{E6}
\end{equation}
\begin{figure}[!htb]
\centering \includegraphics[height=100mm,width=180mm]{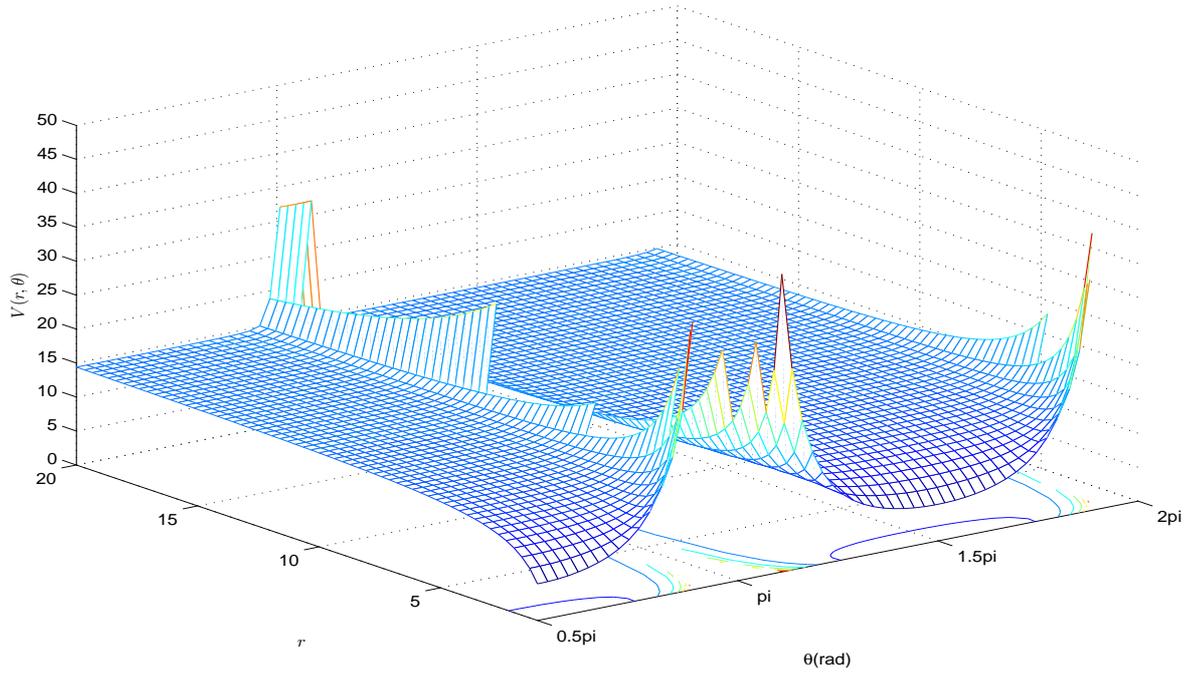}
\caption{{\protect\footnotesize Shape of ring shaped modified Kratzer potential for different values of $r$ and $\theta$.}}
\label{fig1}
\end{figure}
\begin{figure}[!htb]
\centering \includegraphics[height=100mm,width=180mm]{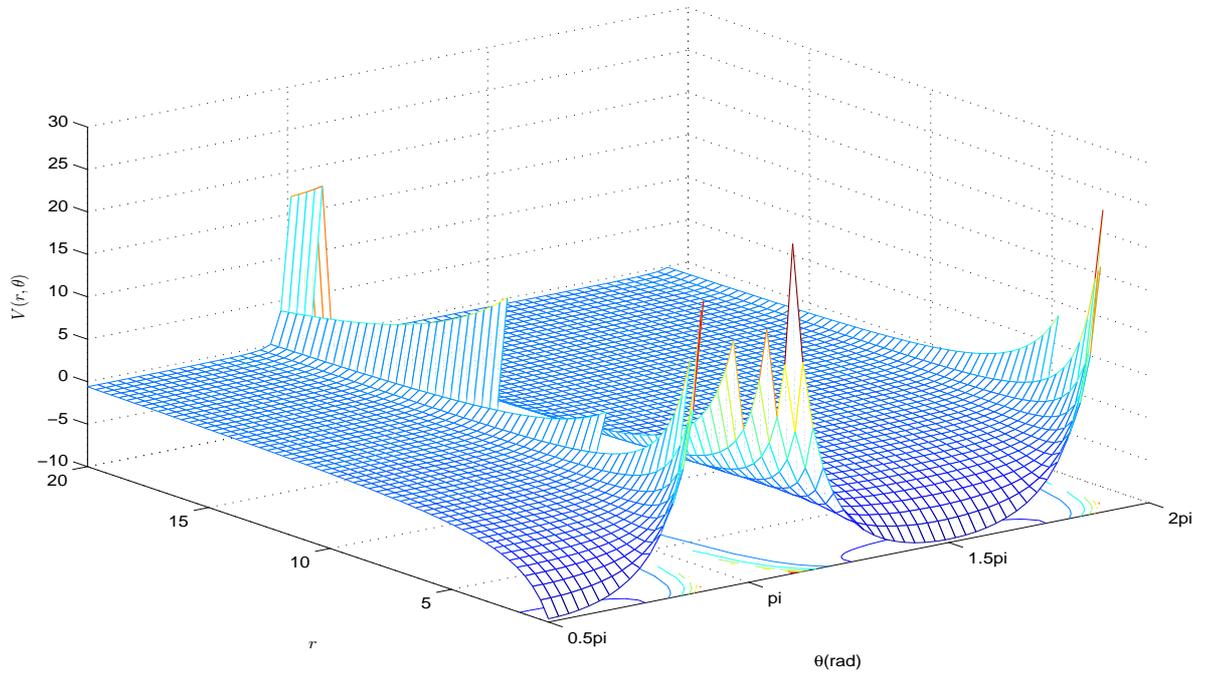}
\caption{{\protect\footnotesize Shape of ring shaped Standard Morse or Kratzer-Fues potential for different values of $r$ and $\theta$.}}
\label{fig2}
\end{figure}
Let us now obtain solution to the angular part of the problem. To perform this task we consider equation (\ref{E4b}) and then introduce a new transformation of the form $\varrho=\cos\theta$ $\in (1, -1)$ which maintained the finiteness of the transformed wavefunctions on the boundary conditions. Thus we have
\begin{equation}
\frac{d^2\Theta(\varrho)}{d\varrho^2}-\frac{2\varrho}{1-\varrho^2}\frac{d\Theta(\varrho)}{d\varrho}+\left[\frac{\ell(\ell+1)-m^2-\varrho^2[\ell(\ell+1)+\eta]}{(1-\varrho^2)^2}\right]=0.
\label{E7}
\end{equation}
The solution to the above equation can be found in terms of the Gegenbauer {polynomials}\footnote{Gegenbauer polynomials or ultraspherical polynomials  are orthogonal polynomials on the interval [-1, 1] with respect to the weight function $(1-x^2)^{\alpha-\frac{1}{2}}$} ${\bf C}_n^m(t)$, which can be referred to as generalization of the Legendre polynomials as:
\begin{equation}
\Theta(\varrho)=2^{-2\zeta} M_{\ell m}(1-z^2)^\zeta {\bf C}_{\tilde{n}}^{2\zeta+\frac{1}{2}}(\varrho)\ \ \ or\ \ \ 
\Theta(\theta)=2^{-2\zeta} M_{\ell m}\sin^\zeta\theta{\bf C}_{\tilde{n}}^{2\zeta+\frac{1}{2}}(\cos\theta),
\label{E8}
\end{equation}
where $\zeta=\sqrt{\frac{m^2+\eta^2}{4}}$ and with the relation
\begin{equation}
\ell+\frac{1}{2}=\left(\sqrt{m^2+\eta}+\tilde{n}+\frac{1}{2}\right).
\label{E9}
\end{equation}
\begin{table}[!htb]
\caption{\scriptsize Spectroscopic parameters and reduced masses for some diatomic molecules composed of a first-row transition metal and main-group elements (H-F).}\vspace*{10pt}{\scriptsize
\begin{tabular}{cccccccc}\hline\hline
{}&{}&{}&{}&{}&{}&{}&{}\\[-1.0ex]
Molecules&$D_e(eV)$&$r_e(A^o)$&$\mu(amu)$&Molecules&$D_e(eV)$&$r_e(A^o)$&$\mu(amu)$ \\[2.5ex]\hline\hline
$ScH$& 2.25& 1.776& 0.986040&$TiC$&2.66& 1.790&9.606079 \\[1ex]
$CrH$ &2.13& 1.694 &0.988976&$NiC$&2.76& 1.621 &9.974265 \\[1ex]
$VH$  &2.33& 1.719 &0.988005&$ScN$&4.56& 1.768&10.682771\\[1ex]
$TiH$& 2.05& 1.781& 0.987371&$ScF$&5.85& 1.794 &13.358942\\[1ex]
$MnH$& 1.67& 1.753& 0.989984&$CuLi$&1.74& 2.310 &6.259494 \\[1ex]
\hline\hline
\end{tabular}\label{tab1}}
\vspace*{-1pt}
\end{table}
\begin{table}[!t]
{\scriptsize
\caption{\normalsize Bound-state energy eigenvalues for ScH, TiH, VH, CrH, MnH molecules for various $n$ and rotational $\ell$ quantum numbers in ring shaped Kratzer-Fues and ring shaped modified Kratzer potential.} \vspace*{10pt}{
\begin{tabular}{cccccccccc}\hline\hline\\[1.5ex]
\multicolumn{4}{c}{}&\multicolumn{2}{c}{ring shaped Kratzer-Fues potential}&\multicolumn{2}{c}{}&\multicolumn{2}{c}{ring shaped modified Kratzer potential}\\[1.5ex]
{}&{}&{}&{}&{}&{}&{}&{}&{}&{}\\[-1.0ex]
Molecules&$n$&$\tilde{n}$&$m$&$\eta=0$&$\eta=10$&&&$\eta=0$&$\eta=10$\\[1ex]\hline\hline
	 	 &0	&0	&0	&0.038550865177		&0.047136152697	&&&-2.2114491348232		&-2.2028638473033	\\[1ex]
	  &1	&1	&0	&0.113671329873		&0.125714585463	&&&-2.1363286701273		&-2.1242854145369	\\[1ex]
ScH	&3	&2	&1	&0.256230689466		&0.268233669609	&&&-1.9937693105341		&-1.9817663303913	\\[1ex]
	  &3	&3	&2	&0.266251159238		&0.278465230223	&&&-1.9837488407615		&-1.9715347697774	\\[1ex]
	  &5	&4	&3	&0.402000580881		&0.412961553329	&&&-1.8479994191190		&-1.8370384466709	\\[1ex]
	  &5	&5	&4	&0.418711228529		&0.429403117956	&&&-1.8312887714706		&-1.8205968820436	\\[1ex]
										
	  &0	&0	&0	&0.036655640099		&0.045168566558	&&&-2.0133443599005		&-2.0048314334419	\\[1ex]
	  &1	&1	&0	&0.108000940853		&0.119913313679	&&&-1.9419990591474		&-1.9300866863210	\\[1ex]
TiH	&3	&2	&1	&0.243080026941		&0.254896743640	&&&-1.8069199730593		&-1.7951032563601	\\[1ex]
	  &3	&3	&2	&0.252945869484		&0.264959557387	&&&-1.7970541305156		&-1.7850404426131	\\[1ex]
	  &5	&4	&3	&0.381216441946		&0.391943093793	&&&-1.6687835580542		&-1.6580569062067	\\[1ex]
	  &5	&5	&4	&0.397565414203		&0.408012370407	&&&-1.6524345857975		&-1.6419876295931	\\[1ex]
										
	  &0	&0	&0	&0.040485742451		&0.049627192146	&&&-2.2895142575488		&-2.2803728078539	\\[1ex]
	  &1	&1	&0	&0.119347125450		&0.132160333302	&&&-2.2106528745497		&-2.1978396666983	\\[1ex]
VH	&3	&2	&1	&0.268892560856		&0.281643439466	&&&-2.0611074391435		&-2.0483565605342	\\[1ex]
	  &3	&3	&2	&0.279537700760		&0.292509080072	&&&-2.0504622992403		&-2.0374909199277	\\[1ex]
	  &5	&4	&3	&0.421812439293		&0.433433943863	&&&-1.9081875607066		&-1.8965660561375	\\[1ex]
	  &5	&5	&4	&0.439528566580		&0.450859092026	&&&-1.8904714334202		&-1.8791409079739	\\[1ex]
										
	  &0	&0	&0	&0.039240240805		&0.048624500805	&&&-2.0907597591951		&-2.0813754991953	\\[1ex]
	  &1	&1	&0	&0.115552366272		&0.128660509524	&&&-2.0144476337276		&-2.0013394904763	\\[1ex]
CrH	&3	&2	&1	&0.259790220955		&0.272748672279	&&&-1.8702097790448		&-1.8572513277214	\\[1ex]
	  &3	&3	&2	&0.270610020743		&0.283775564629	&&&-1.8593899792566		&-1.8462244353706	\\[1ex]
	  &5	&4	&3	&0.407298458878		&0.419010312065	&&&-1.7227015411220		&-1.7109896879354	\\[1ex]
	  &5	&5	&4	&0.425145398908		&0.436538532777	&&&-1.7048546010919		&-1.6934614672227	\\[1ex]
										
	  &0	&0	&0	&0.033530790717		&0.042255441370	&&&-1.6364692092834		&-1.6277445586296	\\[1ex]
	  &1	&1	&0	&0.098573057609		&0.110692944784	&&&-1.5714269423906		&-1.5593070552162	\\[1ex]
MnH	&3	&2	&1	&0.220872295867		&0.232728799779	&&&-1.4491277041331		&-1.4372712002212	\\[1ex]
	  &3	&3	&2	&0.230774127351		&0.242793864141	&&&-1.4392258726494		&-1.4272061358594	\\[1ex]
	  &5	&4	&3	&0.345933039783		&0.356505250760	&&&-1.3240669602166		&-1.3134947492397	\\[1ex]
	  &5	&5	&4	&0.362032877937		&0.372278898839	&&&-1.3079671220633		&-1.2977211011611	\\[1ex]
\hline\hline
\end{tabular}\label{tab2}}
\vspace*{-1pt}}
\end{table}

\begin{table}[!t]
{\scriptsize
\caption{\normalsize Bound-state energy eigenvalues for CuLi, TiC, NiC, ScN and ScF molecules for various $n$ and rotational $\ell$ quantum numbers in ring shaped Kratzer-Fues and ring shaped modified Kratzer potential.} \vspace*{10pt}{
\begin{tabular}{cccccccccc}\hline\hline\\[1.5ex]
\multicolumn{4}{c}{}&\multicolumn{2}{c}{ring shaped Kratzer-Fues potential}&\multicolumn{2}{c}{}&\multicolumn{2}{c}{ring shaped modified Kratzer potential}\\[1.5ex]
{}&{}&{}&{}&{}&{}&{}&{}&{}&{}\\[-1.0ex]
Molecules&$n$&$\tilde{n}$&$m$&$\eta=0$&$\eta=10$&&&$\eta=0$&$\eta=10$\\[1ex]\hline\hline
	&0	&0	&0	&0.0104034334499		&0.0112193128191	&&&-1.7295965665511		&-1.7287806871810	\\[1ex]
	&1	&1	&0	&0.0310236977854		&0.0322097676182	&&&-1.7089763022112		&-1.7077902323822	\\[1ex]
CuLi&3&2	&1	&0.0715256951737		&0.0727926204946	&&&-1.6684743048241		&-1.6672073795052	\\[1ex]
	&3	&3	&2	&0.0725824257022		&0.0738833830266	&&&-1.6674175742921		&-1.6661166169733	\\[1ex]
	&5	&4	&3	&0.1125295661480		&0.1137856501370	&&&-1.6274704338510		&-1.6262143498631	\\[1ex]
	&5	&5	&4	&0.1144499032560		&0.1156950666010	&&&-1.6255500967440		&-1.6243049333992	\\[1ex]
						 				
	&0	&0	&0	&0.0134061298914		&0.0142929020972	&&&-2.6465938701111		&-2.6457070979031	\\[1ex]
	&1	&1	&0	&0.0400156956550		&0.0413085784107	&&&-2.6199843043520		&-2.6186914215897	\\[1ex]
TiC&3	&2	&1	&0.0924312554655		&0.0938202181407	&&&-2.5675687445542		&-2.5661797818594	\\[1ex]
	&3	&3	&2	&0.0935897338979		&0.0950165389951	&&&-2.5664102661090		&-2.5649834610052	\\[1ex]
	&5	&4	&3	&0.1454118148830		&0.1467975683050	&&&-2.5145881851201		&-2.5132024316952	\\[1ex]
	&5	&5	&4	&0.1475306446650		&0.1489052861950	&&&-2.5124693553410		&-2.5110947138057	\\[1ex]
										
	&0	&0	&0	&0.0147961835646		&0.0158370326602	&&&-2.7452038164351		&-2.7441629673411	\\[1ex]
	&1	&1	&0	&0.0441505899397		&0.0456666063198	&&&-2.7158494100600		&-2.7143333936801	\\[1ex]
NiC&3	&2	&1	&0.1019165604800		&0.1035420518290	&&&-2.6580834395201		&-2.6564579481711	\\[1ex]
	&3	&3	&2	&0.1032723340490		&0.1049419108420	&&&-2.6567276659510		&-2.6550580891581	\\[1ex]
	&5	&4	&3	&0.1603378369410		&0.1619561134520	&&&-2.5996621630591		&-2.5980438865482	\\[1ex]
	&5	&5	&4	&0.1628121021450		&0.1644170486870	&&&-2.5971878978550		&-2.5955829513131	\\[1ex]
										
	&0	&0	&0	&0.0168631835648		&0.0176823303543	&&&-4.5431368164352		&-4.5423176696457	\\[1ex]
	&1	&1	&0	&0.0504024684497		&0.0516016960662	&&&-4.5095975315503		&-4.5083983039338	\\[1ex]
ScN&3	&2	&1	&0.1167375787850		&0.1180364898220	&&&-4.4432624212149		&-4.4419635101781	\\[1ex]
	&3	&3	&2	&0.1178209037540		&0.1191557698630	&&&-4.442179096246		&-4.4408442301365	\\[1ex]
	&5	&4	&3	&0.1836207374740		&0.1849280869560	&&&-4.3763792625259		&-4.3750719130435	\\[1ex]
	&5	&5	&4	&0.1856199563870		&0.1869178309770	&&&-4.3743800436129		&-4.3730821690228	\\[1ex]
										
	&0	&0	&0	&0.0168395069607		&0.0174765372110	&&&-5.8331604930393		&-5.8325234627890	\\[1ex]
	&1	&1	&0	&0.0503731014881		&0.0513080511121	&&&-5.7996268985119		&-5.7986919488879	\\[1ex]
ScF&3	&2	&1	&0.1168615557170		&0.1178792669520	&&&-5.7331384442834		&-5.7321207330480	\\[1ex]
	&3	&3	&2	&0.1177103366210		&0.1187564236550	&&&-5.7322896633789		&-5.7312435763445	\\[1ex]
	&5	&4	&3	&0.1837887951430		&0.1848185187380	&&&-5.6662112048568		&-5.6651814812623	\\[1ex]
	&5	&5	&4	&0.1853635624850		&0.1863861908680	&&&-5.6646364375148		&-5.6636138091316	\\[1ex]
	\hline\hline
\end{tabular}\label{tab3}}
\vspace*{-1pt}}
\end{table}

\begin{table}[!t]
{\scriptsize
\caption{\normalsize Fisher's information entropy measures({in unit of $ME^3$}) of ScH, TiH, VH, CrH, MnH, CuLi, TiC, NiC, ScN and ScF molecules for various $n$ and rotational $\ell$ quantum numbers in ring shaped Kratzer-Fues and ring shaped modified Kratzer potential.} \vspace*{10pt}{
\begin{tabular}{cccccccccc}\hline\hline\\[1.5ex]
\multicolumn{4}{c}{}&\multicolumn{2}{c}{Fisher's information entropy}&\multicolumn{2}{c}{}&\multicolumn{2}{c}{Fisher's information entropy}\\[1.5ex]
{}&{}&{}&{}&{}&{}&{}&{}&{}&{}\\[-1.0ex]
Molecules&$n$&$\tilde{n}$&$m$&$\eta=1$&$\eta=10$&&Molecules&$\eta=1$&$\eta=10$\\[1ex]\hline\hline
	  &0	&0	&0	&-0.681899887587	&-0.213093419123	&&	    &-4.430031245300	&-1.398937573470	\\[1ex]
	  &1	&1	&0	&-0.607144881351	&-0.188917645709	&&	    &-4.259482255690	&-1.344365852550	\\[1ex]
ScH	&3	&2	&1	&-0.341956148619	&-0.143113454286	&&	    &-2.785981994730	&-1.185236405610	\\[1ex]
	  &3	&3	&2	&-0.212470375401	&-0.124510654407	&&CuLi	&-1.758191502840	&-1.048191952690	\\[1ex]
	  &5	&4	&3	&-0.118677955570	&-0.084468903166	&&	    &-1.148675272420	&-0.831331357338	\\[1ex]
   	&5	&5	&4	&-0.088094902331	&-0.069901537700	&&	    &-0.877363231883	&-0.707740395896	\\[1ex]
									
	  &0	&0	&0	&-0.574697197357	&-0.179461701496	&&	    &-26.58864033183	&-8.399704277931	\\[1ex]
	  &1	&1	&0	&-0.509851692536	&-0.158488149380	&&	    &-25.72448824362	&-8.123628922161	\\[1ex]
TiH	&3	&2	&1	&-0.285178358925	&-0.119218181398	&&	    &-17.03336998411	&-7.251248387582	\\[1ex]
	  &3	&3	&2	&-0.177004627458	&-0.103606596608	&&TiC	  &-10.75621884151	&-6.417048285671	\\[1ex]
	  &5	&4	&3	&-0.098142306215	&-0.069775317718	&&	    &-7.116418770230	&-5.153923527921	\\[1ex]
	  &5	&5	&4	&-0.072713559835	&-0.057634241562	&&	    &-5.441981377200	&-4.392887202110	\\[1ex]
									
	  &0	&0	&0	&-0.749483394279	&-0.234207036835	&&   	  &-36.22001392321	&-11.44088436361	\\[1ex]
	  &1	&1	&0	&-0.667230790564	&-0.207606407307	&&   	  &-34.96849519112	&-11.04083573651	\\[1ex]
VH	&3	&2	&1	&-0.375702001165	&-0.157230198615	&&	    &-23.05729723281	&-9.813600035213	\\[1ex]
	  &3	&3	&2	&-0.233428975391	&-0.136786905508	&&NiC	  &-14.55726663640	&-8.682768966150	\\[1ex]
	  &5	&4	&3	&-0.130349292522	&-0.092772223552	&&	    &-9.589907385391	&-6.943748569551	\\[1ex]
	  &5	&5	&4	&-0.096751899828	&-0.076767661725	&&	    &-7.330673924481	&-5.916165955761	\\[1ex]
									
	  &0	&0	&0	&-0.673218403489	&-0.210063874680	&&	    &-72.03797020391	&-22.76821398991	\\[1ex]
	  &1	&1	&0	&-0.595029626435	&-0.184772322300	&&	    &-70.31634363684	&-22.21952806162	\\[1ex]
CrH	&3	&2	&1	&-0.330448677958	&-0.137979874288	&&	    &-47.38281312491	&-20.18635563561	\\[1ex]
	  &3	&3	&2	&-0.204872456584	&-0.119770532673	&&ScN	  &-29.94252640000	&-17.87749965170	\\[1ex]
	  &5	&4	&3	&-0.11273096192o	&-0.080053272878	&&	    &-20.16790632931	&-14.61771561520	\\[1ex]
	  &5	&5	&4	&-0.0833551495839	&-0.065993263101	&&	    &-15.44340004458	&-12.47604964143	\\[1ex]
									
	  &0	&0	&0	&-0.434698318885	&-0.135297550767	&&	    &-142.5239497331	&-45.05541946543	\\[1ex]
	  &1	&1	&0	&-0.379837003305	&-0.117550623776	&&  	  &-139.8691771984	&-44.21067895437	\\[1ex]
MnH	&3	&2	&1	&-0.206377377633	&-0.085845284406	&&	    &-95.26420372882	&-40.59913277345	\\[1ex]
	  &3	&3	&2	&-0.127487339584	&-0.074234730389	&&ScF	  &-60.21986000100	&-35.96799843241	\\[1ex]
	  &5	&4	&3	&-0.068529861587	&-0.048482256122	&&	    &-41.00435950272	&-29.73084478700	\\[1ex]
	  &5	&5	&4	&-0.050348153433	&-0.039715438520	&&	    &-31.41838634100	&-25.39078794301	\\[1ex]
\hline\hline
\end{tabular}\label{tab4}}
\vspace*{-1pt}}
\end{table}
\begin{figure}[!t]
\centering \includegraphics[height=100mm,width=180mm]{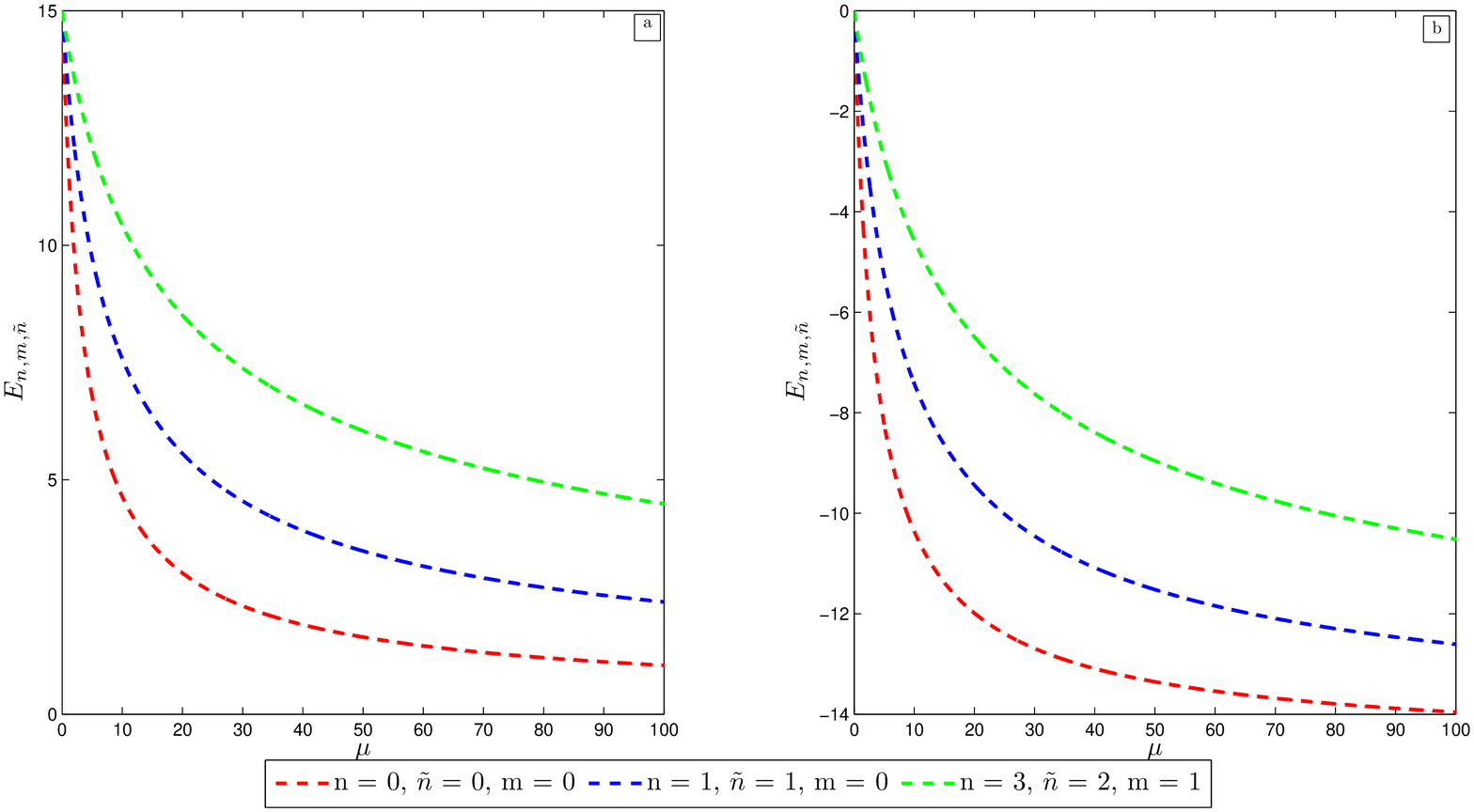}
\caption{{\protect\small (a) The variation of the energy state ($n=0$, $\tilde{n}=0$, $m=0$), ($n=1$, $\tilde{n}=1$, $m=0$) and ($n=3$, $\tilde{n}=2$, $m=1$) of ring shaped modified Kratzer potential as a function of the reduced mass $\mu$. We choose $D_e=15$ $r_e=0.8$, $\eta=10$ and $\hbar=1$. (b) The variation of the energy state ring shaped Kratzer-Fuels as a function of the reduced mass $\mu$}}
\label{fig3}
\end{figure}

\begin{figure}[!t]
\centering\includegraphics[height=100mm,width=180mm]{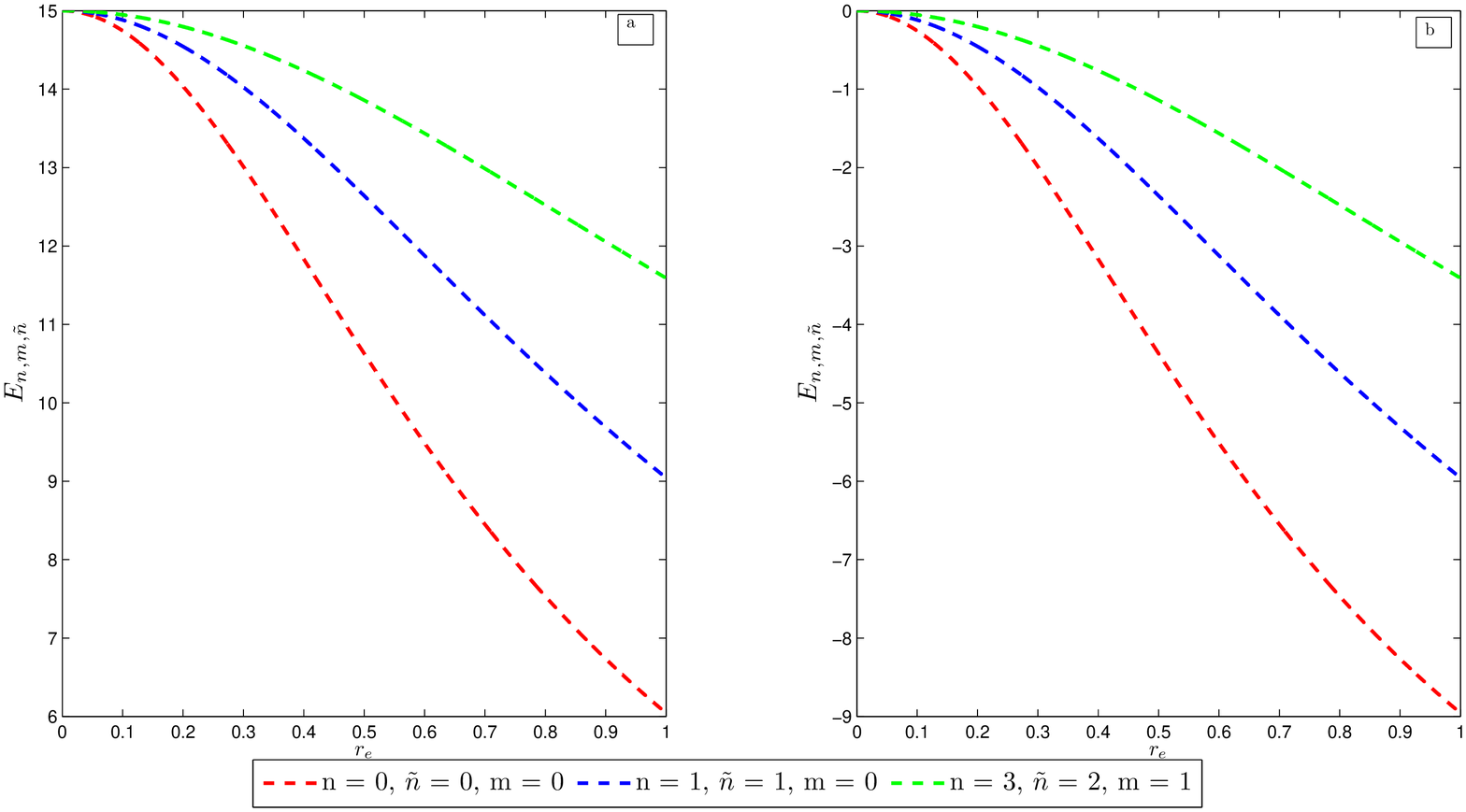}
\caption{{\protect\small (a) The variation of the energy state ($n=0$, $\tilde{n}=0$, $m=0$), ($n=1$, $\tilde{n}=1$, $m=0$) and ($n=3$, $\tilde{n}=2$, $m=1$) of ring shaped modified Kratzer potential as a function of the the equilibrium intermolecular separation $r_e$. We choose $D_e=15$, $\mu=1$; $\eta=10$ and $\hbar=1$. (b) The variation of the energy state ring shaped Kratzer-Fuels as a function of the the equilibrium intermolecular separation $r_e$}}
\label{fig4}
\end{figure}

\begin{figure}[!t]
\centering\includegraphics[height=100mm,width=180mm]{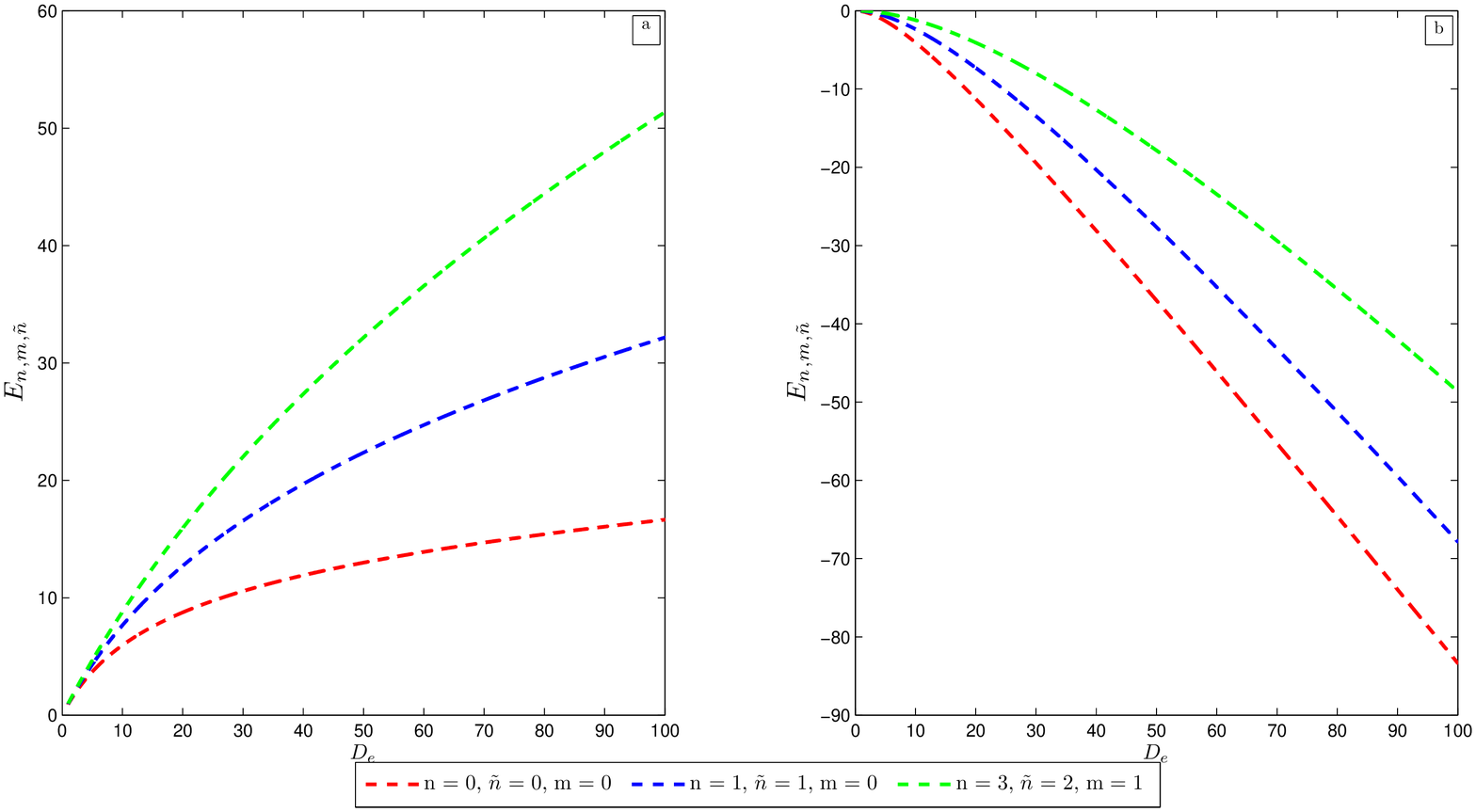}
\caption{{\protect\small (a) The variation of the energy state ($n=0$, $\tilde{n}=0$, $m=0$), ($n=1$, $\tilde{n}=1$, $m=0$) and ($n=3$, $\tilde{n}=2$, $m=1$) of ring shaped modified Kratzer potential as a function of the dissociation energy $D_e$. We choose $\mu=1$, $r_e=0.8$; $\eta=10$ and $\hbar=1$. (b) The variation of the energy state ring shaped Kratzer-Fuels as a function of the dissociation energy $D_e$}}
\label{fig5}
\end{figure}
\begin{figure}[!t]
\centering \includegraphics[height=100mm,width=180mm]{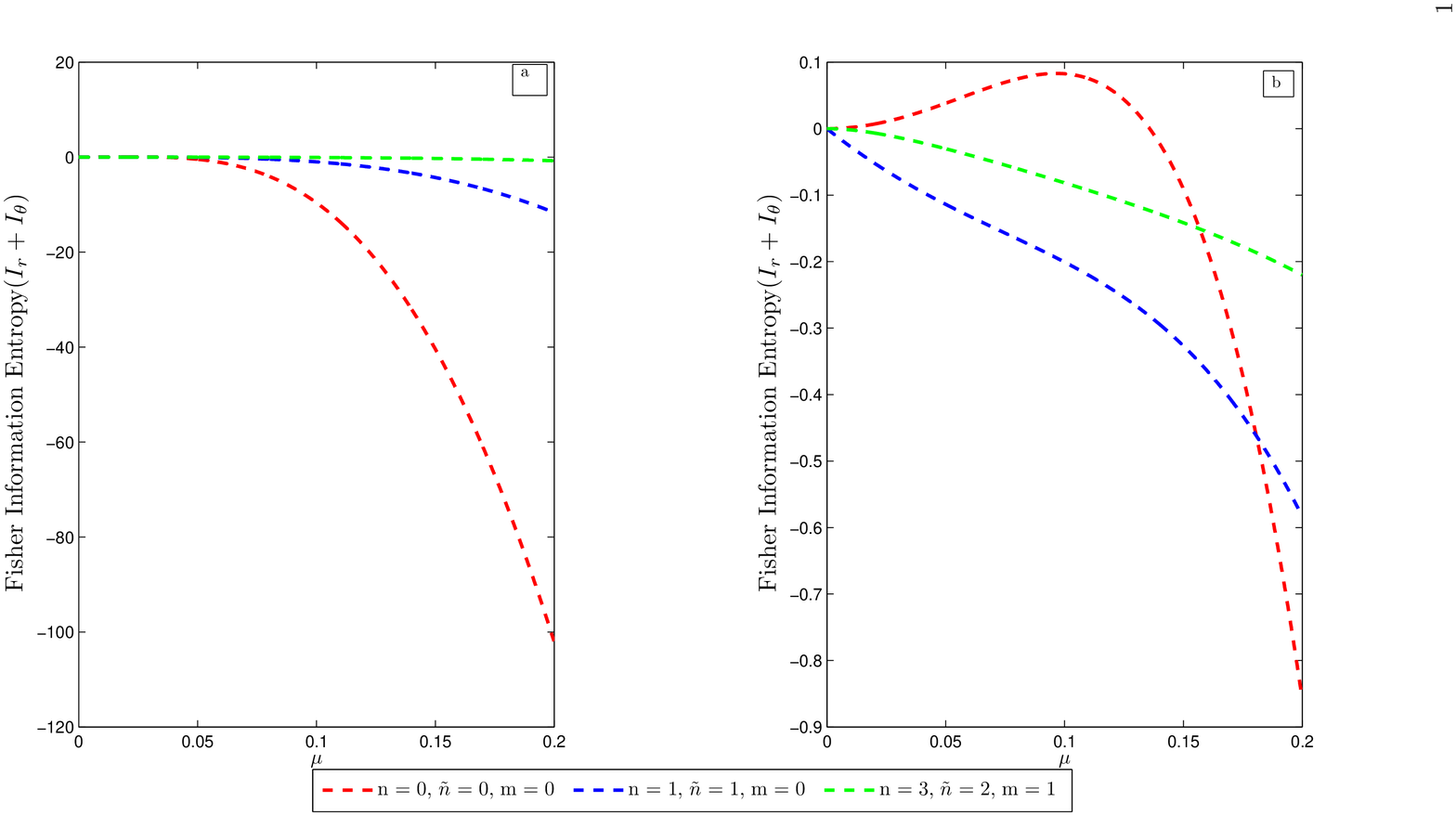}
\caption{{\protect\small (a) The variation of the Fisher's information entropy states ($n=0$, $\tilde{n}=0$, $m=0$), ($n=1$, $\tilde{n}=1$, $m=0$) and ($n=3$, $\tilde{n}=2$, $m=1$) of ring shaped modified Kratzer and ring shaped Kratzer-Fuels potentials as a function of the reduced mass $\mu$. We choose $D_e=15$ $r_e=0.8$, $\eta=1$ and $\hbar=1$. (b) Same as (a) but for $\eta=10$}}
\label{fig6}
\end{figure}

\begin{figure}[!t]
\centering\includegraphics[height=100mm,width=180mm]{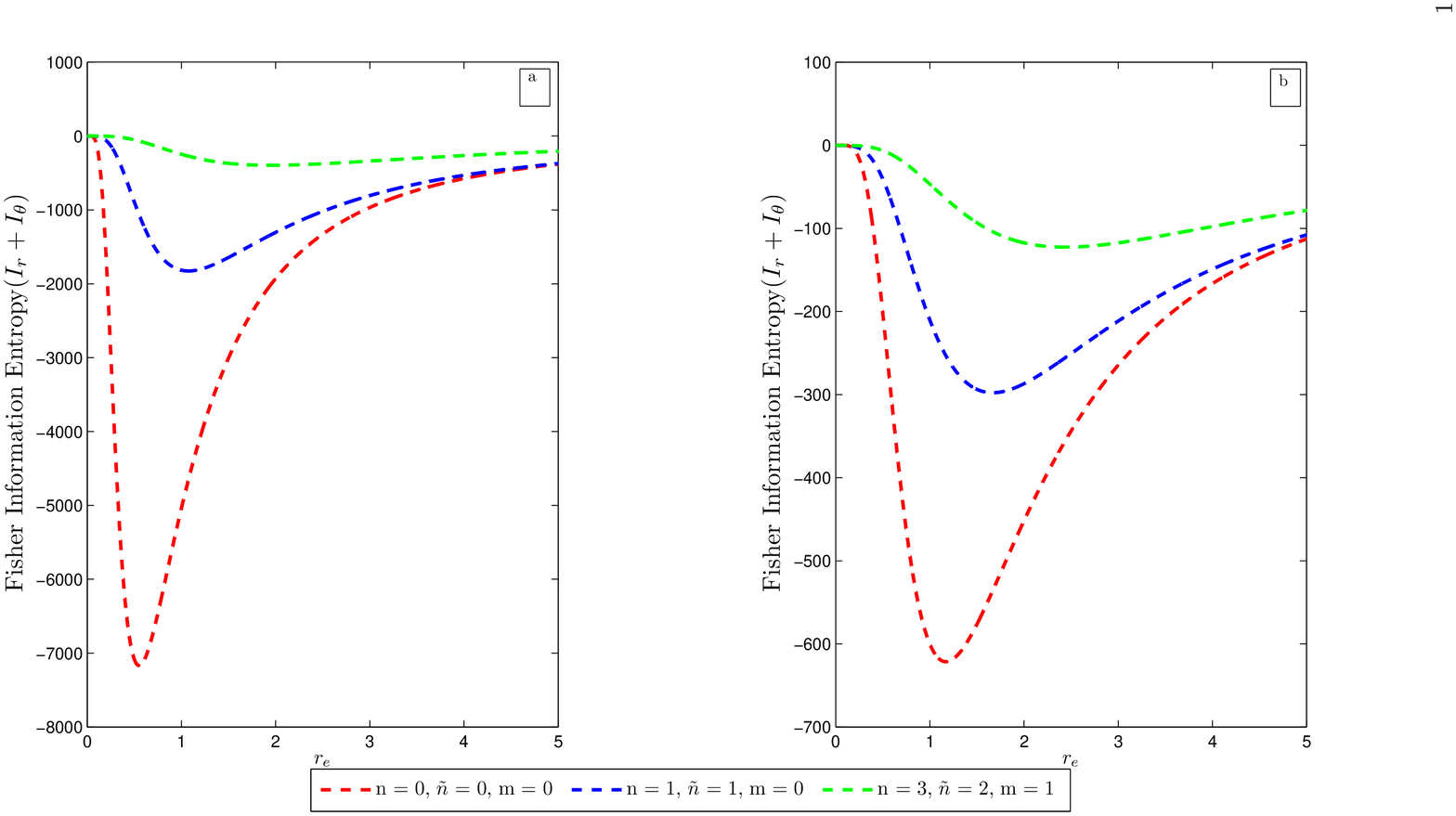}
\caption{{\protect\small (a) The variation of the Fisher's information entropy states ($n=0$, $\tilde{n}=0$, $m=0$), ($n=1$, $\tilde{n}=1$, $m=0$) and ($n=3$, $\tilde{n}=2$, $m=1$) of ring shaped modified Kratzer and ring shaped Kratzer-Fuels potentials as a function of the equilibrium intermolecular separation $r_e$. We choose $D_e=15$ $\mu=1$, $\eta=1$ and $\hbar=1$. (b) Same as (a) but for $\eta=10$}}
\label{fig7}
\end{figure}

\begin{figure}[!t]
\centering\includegraphics[height=100mm,width=180mm]{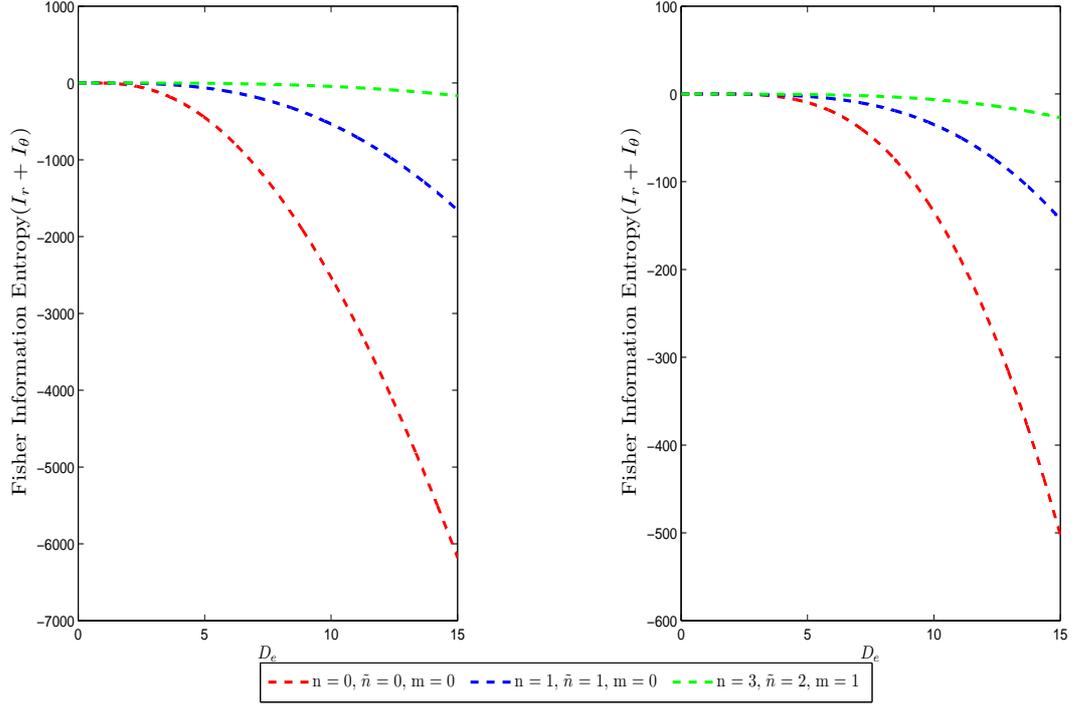}
\caption{{\protect\small (a) The variation of the Fisher's information entropy states ($n=0$, $\tilde{n}=0$, $m=0$), ($n=1$, $\tilde{n}=1$, $m=0$') and ($n=3$, $\tilde{n}=2$, $m=1$) of ring shaped modified Kratzer and ring shaped Kratzer-Fuels potentials as a function of the the dissociation energy $D_e$. We choose $r_e=0.8$ $\mu=1$, $\eta=1$ and $\hbar=1$. (b) Same as (a) but for $\eta=10$}}
\label{fig8}
\end{figure}
Therefore, our final energy levels and eigenfunctions for a real bound charged particle in a Mie field potential plus a combination of non-central potentials can be written as:
\begin{equation}
E_{nm\tilde{n}}=c-\frac{2\mu b^2}{\hbar^2}\left[1+2n+2\sqrt{\left(\sqrt{m^2+\eta}+\tilde{n}+\frac{1}{2}\right)^2+\frac{2\mu a}{\hbar^2}}\right]^{-2}
\label{E10}
\end{equation}
and
\begin{equation}
\Psi_{n\ell m}(r,\theta,\varphi)=2^{-2\zeta}M_{\ell m}N_{n\ell}(\sin\theta)^{2\zeta}\ r^\gamma e^{-\varsigma r}\ {\bf C}_{n}^{2\zeta+\frac{1}{2}}(\cos\theta) L_{n}^{2\gamma+1}(2\varsigma r)e^{\pm im\varphi}.
\label{E11}
\end{equation}
The above equation (\ref{E11}) is normalizable and, thus the normalization constant $N_{n\ell}M_{\ell m}$ can be found by using the following integral formula \cite{N1,BJ20,BJ21}
\begin{equation}
\int_0^\infty\int_0^{\pi}\int_0^{2\pi}\left|\Psi_{n\ell m}(r,\theta,\varphi)\right|^2r^2\sin\theta drd\theta d\varphi=1,
\label{E12}
\end{equation}
through which we found
\begin{equation}
N_{n\ell}M_{\ell m}=\sqrt{\frac{n!(2\varsigma)^{2\gamma+3}}{2(n+\gamma+1)(n+2\gamma+1)!}}\sqrt{\frac{\tilde{n}!\left(\tilde{n}+2\zeta+\frac{1}{2}\right)\left[\Gamma\left(2\zeta+\frac{1}{2}\right)\right]^2}{2\pi^2 2^{-8\zeta}\Gamma(4\zeta+n+1)}},
\label{E13}
\end{equation}
where we have utilized the following standard integrals
\begin{equation}
\int_{-1}^1(1-x^2)^{v-\frac{1}{2}}\left[C_n^v(x)\right]^2dx=\frac{\pi 2^{1-2v}\Gamma(2v+n)}{n!(n+v)[\Gamma(v)]^2}\ \ \mbox{and}\ \ \int_0^\infty x^{\alpha+1}e^{-x}\left[L_n^\alpha\right]^2dx=\frac{(\alpha+n)!(2n+\alpha+1)}{n!}
\label{E14}
\end{equation}
and also we have introduce parameters $\varsigma$ and $\gamma$ in order to avoid mathematical complexity. This parameters are defined as follows:
\begin{equation}
\gamma=-\frac{1}{2}+\sqrt{\left(\ell+\frac{1}{2}\right)^2+\frac{2\mu a}{\hbar^2}} \ \ \ \mbox{and}\ \ \ \ \varsigma=\sqrt{\frac{2\mu}{\hbar^2}(c-E_{n\ell})}.
\label{E15}
\end{equation}
The probability distribution density of this system which gives the probability density for observing the electron in the state characterized by the quantum numbers $(n, l, m)$ is given by
\begin{equation}
\rho(r,\theta,\varphi)=2^{-4\zeta}M_{\ell m}^2N_{n\ell}^2(1-\sin^2\theta)^{2\zeta}\ r^{2\gamma} e^{-2\varsigma r}\ \left[{\bf C}_{\tilde{n}}^{2\zeta+\frac{1}{2}}(\cos\theta)\right]^2 \left[L_{n}^{2\gamma+1}(2\varsigma r)\right]^2.
\label{E16}
\end{equation}
\section{Information-theoretic measures for the Mie-type ring shaped diatomic molecular potential}
In this section, the probability distributions which characterize the quantum-mechanical states of the Mie-type ring shaped diatomic molecular potential are analyzed by means of a complementary information measures of a probability distribution called as the Fisher's information entropy. 

This information measures was originally introduced in 1925 by R. A. Fisher in the theory of statistical estimation \cite{BJ14, BJ15}. This provides the main theoretic tool of the extreme physical information principle and a general variational principle which allows one to derive numerous fundamental equations of physics such as: The Maxwell equations, the Einstein field equations, the Dirac and Klein-Gordon equations, various laws of statistical physics and some laws governing nearly incompressible turbulent fluid flows \cite{BJ16, BJ17, BJ18, BJ19}. This probability distribution which is a gradient functional of their quantum mechanical probability density, is defined as
\begin{equation}
I(\rho):=\int_0^\infty\int_0^{\pi}\int_0^{2\pi}\frac{\left[\bar{\nabla}\rho(r,\theta,\varphi)\right]^2}{\rho(r,\theta,\varphi)}r^2\sin\theta drd\theta d\varphi,
\label{E17}
\end{equation}
where the gradient operator in polar coordinates is given by $\bar{\nabla}=\left(\frac{\partial}{\partial r}, \frac{1}{r}\frac{\partial}{\partial \theta}, \frac{1}{r\sin\theta}\frac{\partial}{\partial \varphi}\right)$. Thus the above equation can be re-written as
\begin{equation}
I(\rho):=\int_0^\infty\int_0^{\pi}\int_0^{2\pi}\frac{1}{\rho(r,\theta,\varphi)}\left[\frac{1}{r}\frac{\partial\rho(r,\theta,\varphi)}{\partial \theta}\right]^2r^2drd\theta d\varphi+\int_0^\infty\int_0^{\pi}\int_0^{2\pi}\frac{1}{\rho(r,\theta,\varphi)}\left[\frac{\partial \rho(r,\theta,\varphi)}{\partial r}\right] r^2drd\theta d\varphi.
\label{E18}
\end{equation}
In order to avoid mathematical complexity, the above integral can be splited into two parts, say $I_\theta$ for the first part and $ I_r$ for the second part. Thus for the first part, the differential of the probability density for observing the electron with respect to $\theta$ can be found as
\begin{equation}
\frac{\partial\rho(r,\theta,\varphi)}{\partial \theta}=\frac{M_{\ell m}^2R_{n\ell}^2(\sin\theta)^{4\zeta}}{2^{4\zeta-1}}\left\{\left({2\zeta+\tilde{n}}\right){\coth\theta}\left[{\bf C}_{\tilde{n}}^{2\zeta+\frac{1}{2}}(\cos\theta)\right]^2-\frac{4\zeta+\tilde{n}}{\sin\theta}\left[{\bf C}_{\tilde{n}-1}^{2\zeta+\frac{1}{2}}(\cos\theta)\right]\left[{\bf C}_{\tilde{n}}^{2\zeta+\frac{1}{2}}(\cos\theta)\right]\right\},
\label{E19}
\end{equation}
where we have used a relations of Gegenbauer polynomials which we derive as
\begin{equation}
\frac{d}{dx}\left\{(1-x^2)^{\lambda-\frac{1}{2}}\left[{\bf C}_n^\lambda(x)\right]^2\right\}=\frac{2\lambda+n-1}{1-z^2}\left[{\bf C}_n^\lambda(x)\right]\left[{\bf C}_{n-1}^\lambda(x)\right]-\frac{\left(\lambda+n-\frac{1}{2}\right)x}{1-z^2}\left[{\bf C}_n^\lambda(x)\right]^2.
\label{E20}
\end{equation}
Thus with equation (\ref{E19}), we can write the first part of the integral as
\begin{equation}
I_\theta(\rho)=\frac{\pi M^2}{2^{4\zeta-3}}\int_0^\infty R(r)^2dr\int_0^\pi(\sin\theta)^{4\zeta+1}\left\{\begin{matrix}\frac{(2\zeta+\tilde{n})^2\cos^2\theta}{\sin^2\theta}\left[{\bf C}_{\tilde{n}}^{2\zeta+\frac{1}{2}}(\cos\theta)\right]^2+\frac{(4\zeta+\tilde{n})^2}{\sin^2\theta}\left[{\bf C}_{\tilde{n}-1}^{2\zeta+\frac{1}{2}}(\cos\theta)\right]^2 \\
-(2\zeta+\tilde{n})(4\zeta+\tilde{n})\frac{\cos\theta}{\sin\theta}\left[{\bf C}_{\tilde{n}-1}^{2\zeta+\frac{1}{2}}(\cos\theta)\right]\left[{\bf C}_{\tilde{n}}^{2\zeta+\frac{1}{2}}(\cos\theta)\right]\end{matrix}\right\}d\theta.
\label{E21}
\end{equation}
In order to solve the integral (\ref{E21}), we decompose it into sum of three integrals says $I_{\theta 1}$, $I_{\theta 2}$ and $I_{\theta 3}$, such that $I_\theta(\rho)=I_{\theta 1}+I_{\theta 2}+I_{\theta 3}$. Thus, by making a variable transformation of the form $x=\cos\theta$, $I_{\theta 1}$ can be calculated as
\begin{eqnarray}
I_{\theta 1}&=&\frac{\pi M^2(2\zeta+\tilde{n})^2\varsigma^2 16^{1-\zeta}}{(n+\gamma+1)(2\gamma+n+1)}\left[\int_{-1}^1(1-x^2)^{2\zeta-1}\left[{\bf C}_{\tilde{n}}^{2\zeta+\frac{1}{2}}(x)\right]^2dx-\int_{-1}^1(1-x^2)^{2\zeta}\left[{\bf C}_{\tilde{n}}^{2\zeta+\frac{1}{2}}(x)\right]^2dx\right]\nonumber\\
           &=&-2\varsigma^2\frac{(2\zeta+\tilde{n})^2(8\zeta+2\tilde{n}+1)}{\zeta(n+\gamma+1)(n+2\gamma+1)},
\label{E22}
\end{eqnarray}
where we have used standard integral (\ref{E14}) and \cite{N1,BJ20,BJ21}
\begin{equation}
\int_{-1}^1(1-x^2)^{v-\frac{3}{2}}\left[C_n^v(x)\right]^2dx=\frac{\pi\Gamma(2v+n)2^{1-2v}}{n!\left(\frac{1}{2}-v\right)\left[\Gamma(v)\right]^2}, \ \ \ \ \frac{x^2}{1-x^2}=\frac{1}{1-x^2}-1.
\label{E23}
\end{equation}
Also, $I_{\theta 2}$ can be calculated as
\begin{eqnarray}
I_{\theta 2}&=&\frac{\pi M^2}{2^{4\zeta-3}}\int_{0}^\infty R(r)^2dr\int_{-1}^1(4\zeta+\tilde{n})^2(1-x^2)^{2\zeta-1}\left[{\bf C}_{\tilde{n}}^{2\zeta+\frac{1}{2}}(x)\right]^2dx\nonumber\\
&=&-2\tilde{n}\varsigma^2\frac{(2\tilde{n}+4\zeta+1)(4\zeta+\tilde{n})}{\zeta(n+\gamma+1)(n+2\gamma+1)}.
\label{E24}
\end{eqnarray}
And finally, $I_{\theta 3}$ reads
\begin{eqnarray}
I_{\theta 3}&=&-\frac{\pi M^2}{2^{4\zeta-3}}\int_{0}^{\infty} R(r)^2dr\int_{-1}^1(4\zeta+\tilde{n})(2\zeta+\tilde{n})(1-x^2)^{2\zeta-1}x\left[{\bf C}_{\tilde{n}}^{2\zeta+\frac{1}{2}}(x)\right]\left[{\bf C}_{\tilde{n-1}}^{2\zeta+\frac{1}{2}}(x)\right]dx\nonumber\\
&=&-\frac{\pi M^2\tilde{n}}{(4\zeta+\tilde{n}-1)2^{4\zeta-3}}(4\zeta+\tilde{n})(2\zeta+\tilde{n})\int_{0}^{\infty} R(r)^2dr\int_{-1}^1{(1-x^2)^{2\zeta-1}\left[{\bf C}_{\tilde{n}}^{2\zeta+\frac{1}{2}}(x)\right]^2}dx\\& &-\frac{\pi M^2(4\zeta+\tilde{n}-1)}{(4\zeta+\tilde{n}-1)2^{4\zeta-3}}(4\zeta+\tilde{n})(2\zeta+\tilde{n})\int_{0}^{\infty} R(r)^2dr \int_{-1}^1{(1-x^2)^{2\zeta-1}\left[{\bf C}_{\tilde{n}}^{2\zeta+\frac{1}{2}}(x)\right]\left[{\bf C}_{\tilde{n}-1}^{2\zeta+\frac{1}{2}}(x)\right]}dx\nonumber\\
&=&2\tilde{n}\varsigma^2\frac{(2\tilde{n}+4\zeta+1)(4\zeta+\tilde{n})(2\zeta+\tilde{n})}{\zeta(4\zeta+\tilde{n}-1)(n+\gamma+1)(n+2\gamma+1)},\nonumber
\label{E25}
\end{eqnarray}
where we have used the following properties of the Gegenbauer polynomials
\begin{equation}
(n+2)\left[{\bf C}_{n+2}^\lambda(x)\right]=2(\lambda+n+1)\left[{\bf C}_{n+1}^\lambda(x)\right]-(2\lambda+n)\left[{\bf C}_n^\lambda(x)\right].
\label{E26}
\end{equation}
By combining $I_{\theta 1}$, $I_{\theta 2}$ and $I_{\theta 3}$ calculated values, we find that
\begin{equation}
I_\theta(\rho)=\frac{2\varsigma^2}{\zeta(n+\gamma+1)(n+2\gamma+1)}\left[\frac{\tilde{n}(4\zeta+\tilde{n})(1-2\zeta)(4\zeta+2\tilde{n}+1)}{4\zeta+\tilde{n}-1}-(2\zeta+\tilde{n})^2(2\tilde{n}+8\zeta+1)\right], 
\label{E27}
\end{equation}
Again, for the second part of the integral($I_r$), the differential of the probability density for observing the electron with respect to $r$ can be found as
\begin{eqnarray}
\frac{\partial \rho(r,\theta,\varphi)}{\partial r}&=&N_{n\ell}^2\left|Y_{\ell m}(\theta, \varphi)\right|^2r^{2\gamma}e^{-2\varsigma r}\left\{2\left(\frac{\varsigma}{r}-1\right)\left[L_n^{2\gamma+1}(2\varsigma r)\right]^2-4\varsigma L_n^{2\gamma+1}(2\varsigma r)L_{n-1}^{2\gamma+2}(2\varsigma r)\right\}\ \ \ \ \nonumber\\
\left[\frac{\partial \rho(r,\theta,\varphi)}{\partial r}\right]^2&=&\rho(r,\theta,\varphi)N_{n\ell}^2\left|Y_{\ell m}(\theta, \varphi)\right|^2r^{2\gamma}e^{-2\varsigma r}\left\{16\varsigma\left(\varsigma-\frac{\gamma}{r}\right)\left[L_n^{2\gamma+1}(2\varsigma r)\right]\left[L_{n-1}^{2\gamma+2}(2\varsigma r)\right]\right.\nonumber\\
&&+\left.4\left(\frac{\gamma}{r}-\varsigma\right)^2\left[L_n^{2\gamma+1}(2\varsigma r)\right]^2+16\varsigma^2\left[L_{n-1}^{2\gamma+1}(2\varsigma r)\right]^2\right\},
\label{E28}
\end{eqnarray}
where relations of the associated Laguerre functions $\frac{d}{dx}L_n^a(x)=-L_{n-1}^{a+1}(x)$ have been used. Thus by putting equation (\ref{E28}) into the second integral, i.e
\begin{equation}
I_r(\rho)=\int_0^\infty\int_0^{\pi}\int_0^{2\pi}\frac{1}{\rho(r,\theta,\varphi)}\left[\frac{\partial\rho(r,\theta,\varphi)}{\partial r}\right]^2r^2drd\theta d\varphi,
\label{E29}
\end{equation}
and then decompose the resulting integral into sum of five integrals; such that we can write
\begin{equation}
I_r(\rho)=I_{r_1}+I_{r_2}+I_{r_3}+I_{r_4}+I_{r_5}.
\label{E30}
\end{equation}
Thus the integrals can be calculated as follows:
\begin{enumerate}
	\item For $I_{r_1}$
	\begin{eqnarray}
	I_{r_1}&=&16N_{n\ell}^2\varsigma\int_0^\infty\left(\varsigma-\frac{\gamma}{r}\right)r^{2\gamma+2}e^{-2\varsigma r}\left[L_n^{2\gamma+1}(2\varsigma r)\right]\left[L_{n-1}^{2\gamma+2}(2\varsigma r)\right]dr\int_0^\pi\int_0^{2\pi}\left|Y_{\ell m}(\theta, \varphi)\right|^2\sin\theta d\theta d\varphi\nonumber\\
	&=&-8\frac{n\varsigma^2}{n+\gamma+1},
	\label{E31}
\end{eqnarray}
where we have utilized the following standard integral and ladder relationship of the Laguerre polynomials for the product $r\left[L_{n-1}^{2\gamma+2}(2\varsigma r)\right]$:
\begin{align}
&\int_0^\infty x^ae^{-x}\left[L_{n}^{a}(x)\right]\left[L_{m}^{a}(x)\right]dx=\frac{\Gamma(a+n+1)}{\Gamma(n+1)}\delta_{mn},\ \frac{x\left[L_{n}^{a+1}(x)\right]}{(n+a+1)}=\left[L_{n}^{a}(x)\right]-\frac{(n+1)}{(n+a+1)}\left[L_{n+1}^{a+1}(x)\right]\nonumber\\
&\mbox{and} \ \ \int_0^\infty x^ae^{-x}\left[L_{n}^{a-1}(x)\right]\left[L_{m}^{a-1}(x)\right]dx=\frac{(a+2n)\Gamma(a+n)}{\Gamma(n+1)}\delta_{mn} 
\label{E32}
\end{align}
\item For $I_{r_2}$
\begin{eqnarray}
	I_{r_2}&=&4N_{n\ell}^2\varsigma\int_0^\infty r^{2\gamma}e^{-2\varsigma r}\left[L_n^{2\gamma+1}(2\varsigma r)\right]^2dr\int_0^\pi\int_0^{2\pi}\left|Y_{\ell m}(\theta, \varphi)\right|^2\sin\theta d\theta d\varphi\nonumber\\
	&=&8\frac{\gamma^2\varsigma^2}{(n+\gamma+1)(n+2\gamma+1)},
	\label{E33}
\end{eqnarray}
where we have we have used the following standard integral \cite{N1,BJ20,BJ21}.
\begin{equation}
\int_0^\infty x^pe^{-x}\left[L_{n}^{a}(x)\right]\left[L_{m}^{b}(x)\right]dx=\Gamma(p+1)\sum_{r=0}^{min(n,m)}(-1)^{n+m}\left(\begin{matrix}p-a\\n-r\end{matrix}\right)\left(\begin{matrix}p-b\\m-r\end{matrix}\right)\left(\begin{matrix}p+r\\r\end{matrix}\right).
\label{E34}
\end{equation}
\item For $I_{r_3}$
\begin{eqnarray}
	I_{r_3}&=&-8N_{n\ell}^2\varsigma\gamma\int_0^\infty r^{2\gamma+1}e^{-2\varsigma r}\left[L_n^{2\gamma+1}(2\varsigma r)\right]^2dr\int_0^\pi\int_0^{2\pi}\left|Y_{\ell m}(\theta, \varphi)\right|^2\sin\theta d\theta d\varphi\nonumber\\
	&=&-8\frac{\gamma\varsigma^2}{(n+\gamma+1)},
	\label{E35}
\end{eqnarray}
where we have used standard integral of the form (\ref{E32}).
\item For $I_{r_4}$
\begin{eqnarray}
	I_{r_4}&=&4N_{n\ell}^2\varsigma^2\int_0^\infty r^{2\gamma+2}e^{-2\varsigma r}\left[L_n^{2\gamma+1}(2\varsigma r)\right]^2dr\int_0^\pi\int_0^{2\pi}\left|Y_{\ell m}(\theta, \varphi)\right|^2\sin\theta d\theta d\varphi\nonumber\\
	&=&4\varsigma^2,
	\label{E36}
\end{eqnarray}
where we have used standard integral of the form (\ref{E14}).
\item For $I_{r_5}$
\begin{eqnarray}
	I_{r_4}&=&16N_{n\ell}^2\varsigma^2\int_0^\infty r^{2\gamma+2}e^{-2\varsigma r}\left[L_{n-1}^{2\gamma+2}(2\varsigma r)\right]^2dr\int_0^\pi\int_0^{2\pi}\left|Y_{\ell m}(\theta, \varphi)\right|^2\sin\theta d\theta d\varphi\nonumber\\
	&=&8\frac{n\varsigma^2}{n+\gamma+1},
	\label{E37}
\end{eqnarray}
where we have used standard integral of the form (\ref{E32}).
\end{enumerate}
By combining the values of $I_{r_1}$,  $I_{r_2}$,  $I_{r_3}$,  $I_{r_4}$,  $I_{r_5}$, the Fisher's information for the radial part can be found as
\begin{equation}
I_r(\rho)=\frac{4\varsigma^2}{n+\gamma+1}\left[\frac{2\gamma^2}{n+2\gamma+1}+n-\gamma+1\right].
\label{E38}
\end{equation}
By putting together the calculated values for $I_\theta(\rho)$ and $I_r(\rho)$ the probability distributions which characterize the quantum-mechanical states of
the Mie-type ring shaped diatomic molecular potential can be obtain via $I_\theta(\rho)$ + $I_r(\rho)$.

\section{Results and Conclusion}
In Table \ref{tab1}, we present spectroscopic constant which are taken from the recent literature \cite{F10}. The authors of ref. \cite{F10} obtained nonrelativistic energy spectrum for this molecules interacting through the shifted Deng-Fan potential. Furthermore, they obtained thermodynamic properties such as, the vibrational mean $U$, specific heat $C$, Free energy $F$ and entropy $S$ in the classical limit for these molecules interacting through the shifted Deng-Fan potential. Moreover, some other interesting works \cite{BJ22, BJ23, BJ24, BJ25, BJ26,BJ27} have also applied these molecules. The increasing interest for using these molecules is nothing but for the purposes which they serve in various aspect of physical, chemical and related  studies \cite{BJ28,BJ29,BJ30, BJ31}. The first column of the table consist of the first-row transition metal hydrides.

Transition metal hydrides are chemical compounds containing a transition metal bonded to hydrogen. Most transition metals form hydride complexes and some are significant in various catalytic, synthetic reactions and in solid matrix samples for infrared spectroscopic study \cite{BJ28}. On the other hand, the next in the list; the transition metal carbide molecules such as TiC and NiC, represent a very active area of findings \cite{BJ28,H1,H2}. This is due to the desire for a quantitative understanding of their chemical bonds type such as:
\begin{enumerate}
\item Salt-like
\item Covalent compounds
\item Interstitial compounds
\item Intermediate transition metal carbides. 
\end{enumerate}
Furthermore, diatomic scandium nitride molecule ScN  has  excellent physical properties of high temperature stability as well as electronic transport properties, which are typical of transition metal nitride \cite{BJ28,H3}. The scandium fluoride molecule ScF is the best studied transition metal halide and it has been fairly well characterized \cite{BJ28,H4}. Diatomic molecules which consist of transition metal and main group elements are challenging theoretically and computationally, but recent advancements in computational methods have made such molecules more accessible to investigations.

The last in our list, is the diatomic molecule which consist of the transition metal element copper (Cu) and the main group element lithium (Li), which elucidates the nature of the bonding in mixed transition metal lithides \cite{BJ28,H5}. All their spectroscopic parameters have been accurately determined by using ab-initio calculations \cite{BJ28}. For complete list of the parameters presented in Table {\ref{tab1}}, see Ref. \cite{BJ30}. In this study, we computed the energy states for these selected diatomic molecules (via the interaction of Mie-type with the noncentral potential)  for various vibrational $n$ and {rotational}\footnote{It should be noted that $\ell$ is function of $\tilde{n}$ and m, i.e., $\ell=f(\tilde{n}, m)$} $\ell$. The results are presented in Tables \ref{tab2} and \ref{tab3}. We have employed the following conversion $1amu = 931.494028MeV/c^2$ and $\hbar c = 1973.29eV\AA$ \cite{BJ23,BJ24,BJ25} throughout our computations. 

In Tables \ref{tab2} and \ref{tab3}, we present results for ring shaped Kratzer-Fues potential (RSK) and ring shaped modified Kratzer potential (RSM). In this regard, we also study the effect of the noncentral potential (NC) on the results.  It can be clearly seen from the two tables that the presence ($\eta=10$) or absent ($\eta=0$) of NC have no significant effect  on the energy states of the RSM and RSK particles with a very narrow band spectrum (i.e., $E^{\eta=0}_{{\tilde{n},m,n}}-E^{\eta=10}_{\tilde{n},m,n}\approx0.01$). We proceed to computing the Fisher's information entropy for this selected diatomic molecules for various $\tilde{n}$, $m$ and $n$. The results are presented in table (\ref{tab4}). We also study the effect of the potential parameter $\eta$ on the energy spectrum. We notice that there is narrow band gap when  ($\tilde{n}=0$, $m=0$, $n=0$) but a wide band gap when  ($\tilde{n}=5$, $m=4$, $n=5$), i.e., $E^{\eta=1}_{{0,0,0}}-E^{\eta=10}_{0,0,0}<E^{\eta=1}_{{5,4,5}}-E^{\eta=10}_{5,4,5}$. In the table \ref{tab4}, we presented the result for RSK and RSM as a single column. This is a result of no significant difference between the two numerical computations, i.e., $F^{RSK}-F^{RSM}\approx10^{-20}$.

In figures \ref{fig3}, \ref{fig4} and \ref{fig5}, we plotted the variation of energy spectrum for the two forms of Mie ring shaped potential as a function of reduced mass, the equilibrium intermolecular separation and the dissociation energy respectively. It is observed that both RSK and RSM follows the same pattern of variation except for figure \ref{fig3}. This is an effect caused by the addition of the dissociation energy in the radial part of RSM. Furthermore, we show the response of the Fisher's information to changes in $\mu$, $r_e$ and $D_e$ in figures \ref{fig6}, \ref{fig7} and \ref{fig8}. The effect of ring shaped potential have also been displayed graphically.

In summary, we have obtained the nonrelativistic $\ell-$state solutions of Mie-type ring shaped potential. Some numerical computations for various diatomic molecules of interest have been presented. The dependance of the energy eigenvalues on $\mu$, $r_e$ and $D_e$ have also been explained by graphical representation. In order to have the conceptual understanding of the quantum system under consideration, we have analyze our solution by means of a complementary information measures of a probability distribution called as the Fisher's information entropy. The variation of this information entropy on parameters $\mu$, $r_e$ and $D_e$ have also been explained graphically.

\end{changemargin}
\end{document}